\documentclass[%
reprint,superscriptaddress,amsmath,amssymb,pre,aps,twocolumn,
showkeys
,nofootinbib
]{revtex4-2}
\usepackage{dcolumn}
\usepackage{bm}
\usepackage{listings}
\usepackage{ulem}
\usepackage{graphicx}
\usepackage{subfigure}
\newcommand{\veca}{{\vec{a} } }

\newcommand{\vecX}{{\vec{X}}}
\newcommand{\hatS}{{\hat{S}}}
\newcommand{\la}{\langle}
\newcommand{\ra}{\rangle}
\newcommand{\beq}{\begin{equation}}     
\newcommand{\eeq}{\end{equation}}
\newcommand{\beqa}{\begin{eqnarray}}    
\newcommand{\eeqa}{\end{eqnarray}}
\newcommand{\inv}[1]{{\frac{1}{#1}}}
\newcommand{\inAverage}[1]{{\left\la{#1}\right\ra}}
\begin{document}
\title{{Martingale drift of Langevin dynamics and classical canonical spin statistics-II}
}
\author{Ken Sekimoto}
\affiliation{Laboratoire Mati\`ere et Syst\`emes Complexes, UMR CNRS 7057, Universit\'e Paris Cit\'e,\\ 		10 Rue Alice Domon et L\'eonie Duquet, 75013, Paris, France }
\affiliation{Laboratoire Gulliver, UMR CNRS 7083, ESPCI Paris, Universit\'e PSL\\ 		10 rue Vauquelin, 75005, Paris, France.}
\email[Corresponding author: ]{ken.sekimoto@espci.fr}
	
\date{ \today} 

\begin{abstract} 
% pre@aps.org : Correction EV12344 Sekimoto
In the previous paper we have shown analytically that, if the drift function of the d-dimensional Langevin equation is the Langevin function with a properly chosen scale factor, then the evolution of the drift function is a martingale associated with the histories generated by the very Langevin equation. Moreover, we numerically demonstrated that those generated histories from a common initial data become asymptotically ballistic, whose orientations obey the classical canonical spin statistics under the external field corresponding to the initial data. In the present paper we provide with an analytical explanation of the latter numerical finding by introducing a martingale in the spin functional space. In a specific context the present result elucidates a new {\it physical} aspect of martingale theory. 
\end{abstract}
%\\ \indent Key words  :  \\
\maketitle
%%%%%%%
%\tableofcontents
\section{Introduction}
\subsection{Background and outline}
Within the context of stochastic process in physics \cite{Peliti-book}, martingale theory played an important rule since 2010's, 
 having given a unifying viewpoint for the known results of stochastic thermodynamics, and having brought new results through the application of the known theorems of martingale \cite{martingale-Gupta2011,Roldan-prX2017,review300AdvInPhys}. Yet another interest in the martingale is whether this theory has some physical ingredients or consequences in itself. In the previous paper  \cite{PQL2024} (hereafter ``Paper I'') we have given a positive answer.
There we investigated the consequences of the martingality imposed on the drift function of a Langevin equation \cite{langevinpaper} while the very Langevin equation defines the path probability measure for the martingale, the idea which we called ``self-harmonic drift.''
The outcome was that such a drift function is what usually describes the canonical equilibrium polarisation of a classical Heisenberg spin as function of the external magnetic field with proper scaling. In brief, the Langevin equation met with the Langevin function \cite{langevinfunction} through the martingale. In Paper I this result was analytically derived and confirmed numerically. 
Besides, there we have had also a strong numerical evidence that the canonical spin distribution which would give rise to the initial (self-harmonic) drift emerges in the asymptotic long-time trajectories of the Langevin evolution, where each asymptotically ballistic trajectory defines a spin.
The mechanism behind this phenomenon, however, was left as a puzzle in Paper I, and in the present paper we resolve this analytically. The key to the puzzle is that, in the classical spin {\it functional} space, the canonical spin density function realises a martingale process, which we call the functional martingale.

The organisation of the paper is as follows: In \S\S\ref{subsec:setup} we introduce
the setup and summarise the previous results. In \S\ref{sec:functional-martingale} 
we introduce the functional martingale, first showing the central result obtained from it 
 (\S\S\ref{subsec:central-result}), then provide some heuristic arguments to motive this concept (\S\S\ref{subsec:heuristics}). Using these results, we solve in \S\ref{sec:conseq} the puzzle left in Paper I. The last section \S\ref{sec:conclusion} is for conclusion.

\subsection{Setup and resum\'e of previous results}\label{subsec:setup}

To be self-contained, the main results of Paper I \cite{PQL2024} are recapitulated below along with some definitions. We shall denote by $\vecX_{[t_1,t_2]}$  a history of the variable $\vecX$ over the time interval $[t_1,t_2].$ 
Given a probability space of Markov processes $\{\vecX_{[0,\infty]}\}$, the function $\Phi(\vecX)$ is said to be a harmonic function if the induced process $\Phi(\vecX_t)$ is a Martingale process, i.e., 
\beq \label{eq:martingale}
\la \Phi(\vecX_t) |\vecX_s\ra =\Phi(\vecX_s), \quad t\ge s\ge 0
\eeq
holds, where $\la\,|\vecX_{s} \ra$ means the conditional expectation subject to a specified value, $\vecX_{s}.$\footnote{ Martingale are usually defined in terms of conditioning on a sub-history $\{\vecX_{[0,s]}\},$ but  since we assume $\{\vecX_{[0,\infty]}\}$ to be Markovian,
and since $\Phi(\vecX_t)$ is the function of a single time, the single condition like $\la\,|\vecX_s\ra $ suffices.}
As long as the It\^o differentiation and the expectation operation are commutative
\footnote{cf. The so-called the strict local martingale may not assure this; 
see, for example, \cite{review300AdvInPhys} \S 4.2.2}, the above property (\ref{eq:martingale}) implies 
\beq \label{eq:dmartingale}
\la d\Phi(\vecX_t) |\vecX_s\ra =0, \quad t\ge s\ge 0.
\eeq
In particular, when the path probability space of $\{\vecX_{[0,\infty]}\}$ is generated by the following $d$-dimensional Langevin equation (and initial conditions), 
\beq\label{eq:Leq}
d\vecX_t=\veca(\vecX_t)dt +d\vec{W}_t,
\eeq
condition (\ref{eq:dmartingale}) is satisfied if 
$$ 
\mathcal{L}_{a^{}}^\dag \Phi(\vecX)=0, 
$$
where $\mathcal{L}_{a^{}}^\dag$ is the adjoint of the Fokker-Planck operator associated with  (\ref{eq:Leq}): 
\beq \label{eq:Ldag}
\mathcal{L}_{a^{}}^\dag \equiv \veca^{}(\vecX)\!\cdot\!\nabla +\inv{2}\Delta.
\eeq
In fact the above sufficient condition is understood by applying the It\^o's theorem to $d\Phi(\vecX_t)$ and taking the conditional expectation for $t\ge s\ge 0$:
\beq \label{eq:B}
{d}\inAverage{\left.\Phi(\vecX_t)\right|\vecX_s}=\inAverage{\left.\mathcal{L}_{a^{}}^\dag \Phi(\vecX_t) \right|\vecX_s}dt.
\eeq

With this basic setup, the result of the first half of Paper I is the emergence of 
the Langevin function when we impose that the drift function $\veca(\vecX)$ of the Langevin equation (\ref{eq:Leq}) is a martingale (i.e. harmonic) associated with the very Langevin evolution: Denoting such a drift function as $\veca^{\,*}(\vecX)$, this implies 
\beq\label{eq:auto-harm}
\mathcal{L}_{a^{\! *}}^\dag \veca^{\,*}(\vecX)=0.
\eeq
This non-linear equation is satisfied by the equilibrium classical spin response, 
\beq \label{eq:vecas}
\veca^{\,*}(\vecX)=L_d(\|\vecX\|) \,{ \frac{\vecX}{\|\vecX\|}},
\eeq
where $L_d(x)$ is the $d$-dimensional Langevin function, e.g. $L_3(x)=\coth{x}-\inv{x}$ in three dimension. 
In the context of the canonical statistics,  $\veca^{\,*}(\vecX)$ means the equilibrium polarisation of a classical spin $\hatS$ under a non-dimensionalized external field  $\vecX:$ 
\beq \label{eq:astar}  
\veca^{\,*}(\vecX) =
\oint_{\|\hatS\|=1}\hatS \, {\mathcal{P}_S^{\rm (can)}}(\hatS;\vecX) d\Omega_\hatS,
\eeq
where
$\oint_{\|\hatS\|=1} f(\hatS) d\Omega_{\hatS}$
stands for a uniform integral of $f(\hatS)$ over a $d$-dimensional spin, i.e. an integral over the $(d-1)$-dimensional unit hypersphere, and 
 ${\mathcal{P}_S^{\rm (can)}}(\hatS;\vecX)$ is the canonical spin density,
which reads
\beq \label{eq:Pcan}
{\mathcal{P}_S^{\rm (can)}}(\hatS;\vecX)
 \equiv \frac{e^{\hatS\cdot\vecX}}{\oint_{\|\hatS\_\|=1} e^{\hatS\_\cdot\vecX}
  d\Omega_{\hatS\_} }, \quad \|\hatS\|=1.
\eeq 

\begin{figure}[t!!] 
\includegraphics[width = 0.7 \linewidth]{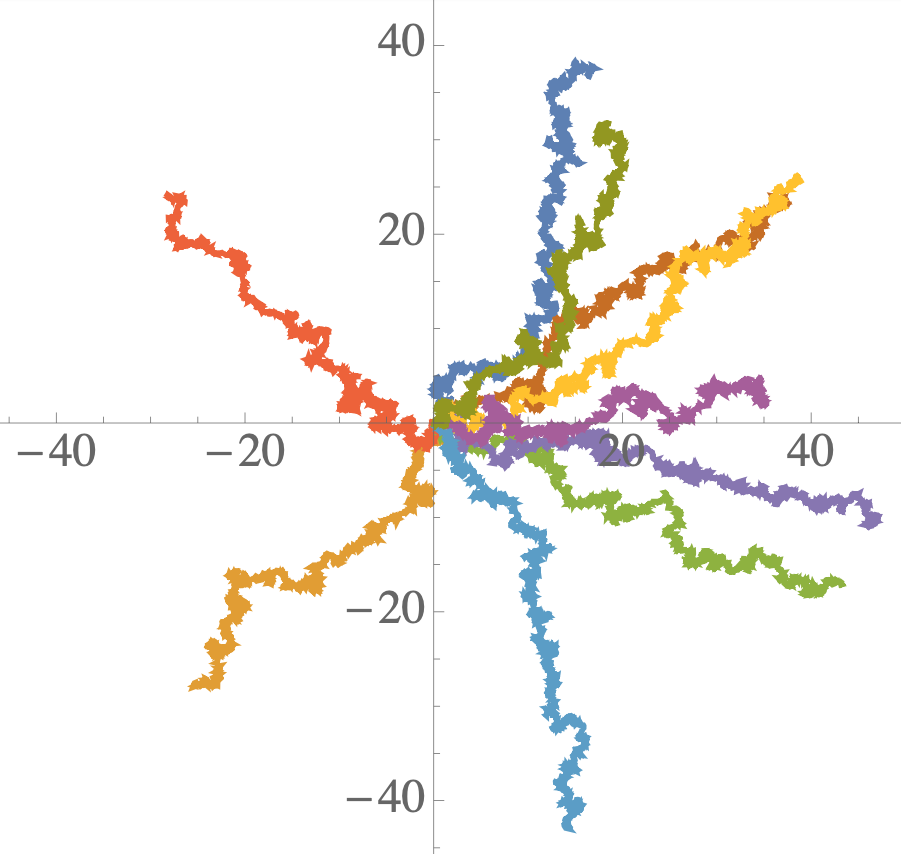}
\caption{ 
Ten trajectories of $\{\vecX_{0\le t \le 40}\}$ all starting from 
$\vecX_0=(1,0)$ and following the Langevin equation (\ref{eq:Leq}) in 2d with the 
drift $\veca^{}(\vecX)$ being given by the self-harmonic drift satisfying (\ref{eq:auto-harm}), more concretely, $\veca^{\,*}(\vecX)=
L_2(\|\vecX\|) (\vecX/\|\vecX\|),$ with $L_2(x)=I_1(x)/I_0(x),$
where $I_n(x)$ ($n=0,1$) are the $n$-th modified Bessel function of the second kind. At $t=40$ a tendency towards ballistic trajectories may be already seen.
(This figure has been modified from Fig.2(b) of Paper I \cite{PQL2024}.)
}\label{fig:trajectories}
\end{figure}

The second half of Paper I reported the numerical observation that, when the process $\vecX_t$ starts from a common initial value $\vecX_0$ and evolves according to (\ref{eq:vecas}), the asymptotic orientation of the trajectory, $(\null{\vecX_t}/{\| \vecX_t\|}),$ at long-time limit obeys statistically as if it were a spin obeying (\ref{eq:Pcan}) with $\vecX=\vecX_0.$
Fig.\ref{fig:trajectories}, which essentially recapitulates Fig.2(b) of Paper I \cite{PQL2024}, shows an example in $d=2$ with $\vecX_0=(1,0),$ and the trajectories up to $t=40$ are drawn in different colours. Up to time $t=40,$ we see already a tendency of converging orientation, (\ref{eq:defas}). 
As in Paper I, we will denote by $\veca^{\,*}_\infty$ this asymptotic orientation of the trajectory. This ``spin'' can be defined by different but equivalent manners:
\beq \label{eq:defas}
\veca^{\,*}_\infty \equiv \lim_{t\to \infty}\frac{\vecX_t}{t} =\lim_{t\to \infty}\frac{\vecX_t}{\| \vecX_t\|}=\lim_{t\to \infty} \veca^{\,*}(\vecX_t),
\eeq
where we have used that $\lim_{x\to \infty}L_d(x)=1.$ 
In any case, the above numerical assertion can be expressed as
\beq \label{eq:tobemet}
{\mathcal{P}^*}(\veca^{\,*}_\infty|\vecX_0)={\mathcal{P}_S^{\rm (can)}}(\hatS=\veca^{\,*}_\infty;\vecX_0),
\eeq
where we denote by ${\mathcal{P}^*}(\veca^{\,*}_\infty|\vecX_0)$ 
 the probability density for the ``spin'' $\veca^{\,*}_\infty$ on the unit hyper-surface. We next give a proof for (\ref{eq:tobemet}).
%%%%%%
%%%%%% 

\section{Functional martingale}\label{sec:functional-martingale} 
\subsection{Central result}\label{subsec:central-result}
The following mathematical identity is the central result in the proof of (\ref{eq:tobemet}):
\beq\label{eq:LdagP}
\mathcal{L}_{a^{\! *}}^\dag
{\mathcal{P}_S^{\rm (can)}}(\hatS;\vecX)=0.
\eeq
Having the definitions (\ref{eq:Ldag}) and (\ref{eq:Pcan}), the verification of (\ref{eq:LdagP}) is straightforward and easy. Based on the argument around (\ref{eq:B}) in \S\ref{subsec:setup} it follows that $\mathcal{P}_S^{\rm (can)} (\hatS;\vecX_t) $ is a martingale process associated with the reference process, $\vecX_t.$
Then the martingale property (\ref{eq:martingale}) reads
\beq \label{eq:mtglP}
\la {\mathcal{P}_S^{\rm (can)}}(\hatS;\vecX_t)
 |X_{s}\ra ={\mathcal{P}_S^{\rm (can)}}(\hat{S};\vecX_s)  
\quad \mbox{for $\forall t\ge s \ge 0$}.
\eeq  
Once we have (\ref{eq:mtglP}), the result that $\veca^{\,*}(\vecX_t)$ is martingale immediately follows from (\ref{eq:astar}). In \S\ref{sec:conseq} we will prove the assertion of Paper I, (\ref{eq:tobemet}),  from (\ref{eq:mtglP}).

While it is easy to verify (\ref{eq:LdagP}), it might look {\it ad hoc} at the same time.
In fact we have not yet a {\it physical} understanding of this result.
Instead, we present below a heuristic argument to motivate (\ref{eq:LdagP}). 
The next subsection \S\ref{subsec:heuristics} is, for the moment, logically disconnected from
the proof of \ref{eq:tobemet}) in \S\ref{sec:conseq}.

\subsection{Heuristic argument for (\ref{eq:LdagP})}\label{subsec:heuristics}
Consider an ensemble of processes $\{X_{[t,\infty]}\}$ starting from the common initial point $\vecX_t$ and henceforth
evolving according to the Langevin equation under self-harmonic drift, i.e., (\ref{eq:Leq}) with $\veca=\veca^{\,*}.$  
We shall focus on ${\mathcal{P}^*}(\veca^{\,*}_\infty |\vecX_t)$ in (\ref{eq:tobemet}), the conditional probability density for 
$\veca^{\,*}_\infty$ with specified value of $\vecX_t,$ where $t$ is regarded as the initial time.
We, however, soon realize that the initial time does not matter because the elapsed time is infinite: We can take advantage of this fact to proceed in the spirit of ``renormalisation group (RG) equation'' {approach to the} singular perturbation problem of partial differential equations {by Goldenfeld and Oono} \cite{ChenGldenOono-RG0,ChenGldenOono-RG,Goldenfeld-book}.
More concretely, we express explicitly the arbitrariness of the initial time as follows:
$$
{\mathcal{P}^*}(\veca^{\,*}_\infty |\vecX_t=\vecX)={\mathcal{P}^*}(\veca^{\,*}_\infty |\vecX_{t-dt}=\vecX),
$$
and for the right hand side we express a convolution form 
\beqa \label{eq:RG}
{\mathcal{P}^*}(\veca^{\,*}_\infty |\vecX_{t-dt}=\vecX)
 &=&
\int_{\vecX'} {\mathcal{P}^*}(\veca^{\,*}_\infty |\vecX_{t}=\vecX')
 \cr && \!\!\!\!\!\!\!\!\!\!\!\!\!\!\!\!\!\!
\times \inAverage{\left.\delta(\vecX'-\hat\vecX_t)\right| \vecX_{t-dt}=\vecX}d \vecX'.
\eeqa
where the second factor of the integrand is the $(\vecX',\vecX)$ element of the 
Markovian propagator from the time $t-dt$ to $t.$ 
Because $\hat\vecX_t$ in the delta-function reads $\vecX+\veca^{\,*}(\vecX)dt+d\vec{W}_{t-dt},$ the development of $\delta$ function in the propagator up to the order $dt$ using the It\^o's theorem allows to rewrite (\ref{eq:RG}) as follows:
\beq \label{eq:PasPcan}
\mathcal{L}_{a^{\! *}}^\dag {\mathcal{P}^*}(\veca^{\,*}_\infty |\vecX_t)=0.
\eeq
Therefore, the unknown {\it density} ${\mathcal{P}^*}(\veca^{\,*}_\infty |\vecX_t)$ 
as an induced process should be martingale.

On the other hand, in Paper I we have numerically demonstrated that the distribution 
of ${a^*_\infty}$ should obey the equilibrium spin distribution of which the external field 
is the initial value of $\vecX,$ which we write $\vecX_t$:
\beq \label{eq:vecascan}
{\mathcal{P}^*}(\veca^{\,*}_\infty |\vecX_t) ={\mathcal{P}_S^{\rm (can)}}(\veca^{\,*}_\infty;\vecX_t),
\quad \mbox{\small (numerical assertion).}
\eeq
In fact we can directly verify that the hypothesis (\ref{eq:vecascan}) 
satisfies  (\ref{eq:PasPcan}) if $\veca^{\,*}_\infty$ has the unitary norm. 
While this verification is apparently in favor of the hypothesis  (\ref{eq:vecascan}), there is a caveat: When we retrace the above argument from the introduction of 
${\mathcal{P}^*}(\veca^{\,*}_\infty |\vecX_t),$ nowhere have we used the fact that $\veca^{\,*}_\infty$ is an asymptotic drift having the unitary norm. Indeed we should rather  replace this $\veca^{\,*}_\infty$ by the generic spin variable $\hatS.$ 
In doing so, we can verify the identity (\ref{eq:LdagP}) from scratch as said before.
The valid derivation of (\ref{eq:tobemet}) will then follow from (\ref{eq:mtglP}) as shown below.

%%%%%%
\section{Consequences of functional martingality}\label{sec:conseq}
We first point out the generic feature of the martingale
(\ref{eq:mtglP})  as process in the functional space. Then we discuss the derivation of (\ref{eq:tobemet}) by way of (\ref{eq:mtglP}).

It is known that the Girsanov martingale, $Y_t=Y_0 e^{b W_t-\frac{b^2}{2}t},$ can generate a series of martingale processes by expanding with respect to $b$ around $b=0$. (See, for example, \S2.2.2 of \cite{review300AdvInPhys} and \cite{Williams1979} cited therein.) In a similar manner, the functional martingale,
$Y_t={\mathcal{P}_S^{\rm (can)}}(\hatS;\vecX_t),$ can generate a series of martingale processes by differentiating by $\hatS$ in the $(d-1)$-dimensional tangent space. 
Let us denote by $\nabla_S$  the gradient operating as if $\hatS$ were $d$-dimensional Euclidean vector. Then the first derivative martingale is such that
 ${\mathcal{P}_S^{\rm (can)}}(\hatS;\vecX_t)$ in  (\ref{eq:mtglP}) is replaced by 
  $(1-\hatS\hatS)\!\cdot\!\vecX_t \, {\mathcal{P}_S^{\rm (can)}}(\hatS;\vecX),$ 
more precisely, by its tangential component discarding the longitudinal dimension along $\hatS$.

Now we will show how the martingale (\ref{eq:mtglP}) implies the property (\ref{eq:tobemet}).
In (\ref{eq:mtglP}) we let $s\downarrow 0$ on the one hand, and let $t\uparrow \infty$ on the other hand.
As was shown in Paper I, the trajectory $\vecX_t$ asymptotically becomes ballistic, meaning that the field strength $\|\vecX_t\|$ is unboundedly strong, whose orientation is $\veca^{\,*}_\infty.$ In this limit the canonical spin density ${\mathcal{P}_S^{\rm (can)}}(\hatS;\vecX_t)$ weakly converges to the delta function;
$$
{\mathcal{P}_S^{\rm (can)}}(\hatS;\vecX_t)
\to \delta(\hat{S},\veca^{\,*}_\infty),
\quad t\to\infty
$$ 
where $ \delta(\hat{S},\hat{S}')$ stands for the delta function on the unit hyper-surface. (See, for example, \S28 of \cite{Kolmogorov1957} about the weak topology.) Note that $\veca^{\,*}_\infty$ %top-modif
introduced in (\ref{eq:defas}) is the random variable of unitary norm 
which is adapted to the path $\vecX_{[0,\infty]}.$ %end-modif
Then (\ref{eq:mtglP}) under the double temporal limit, $(s,t)\to (0,\infty),$ reads
\beq \label{eq:weakeq}
\la \delta(\hat{S},\veca^{\,*}_\infty) |X_{0}\ra 
={\mathcal{P}_S^{\rm (can)}}(\hatS;\vecX_0).
\eeq
As a weak equality, the operation of $\oint_{\|\hatS\|=1}d\Omega_\hatS \psi(\hatS)$ on each side of (\ref{eq:weakeq}) with {\it any} well-behaved function of spin, $\psi(\hatS),$ should yield the equality,
$$ 
\la \psi(\veca^{\,*}_\infty ) |\vecX_0\ra =
\oint_{\|\hatS\|=1} \psi(\hatS) {\mathcal{P}_S^{\rm (can)}}(\hatS;\vecX_0)
d\Omega_\hatS.
$$
The last equality means that $\veca^{\,*}_\infty$ obeys the same statistics as 
the canonical spin $\hatS$ under the field $\vecX_0.$
This is what (\ref{eq:tobemet}) meant, that is,
${\mathcal{P}^*}(\veca^{\,*}_\infty|\vecX_0)={\mathcal{P}_S^{\rm (can)}}(\hatS=\veca^{\,*}_\infty;\vecX_0).$

\section{Conclusion}\label{sec:conclusion}
We proved that the asymptotic ballistic trajectories of $\vecX_t$ following 
a Langevin equation with self-harmonic drift $\veca^{\,*}$ obey canonical spin statistics under the initial field $\vecX_0.$ 
Nevertheless, the appearance of the canonical distribution, ${\mathcal{P}_S^{\rm (can)}}(\hatS;\vecX_t),$ in a martingale context remains enigmatic.  A better understanding of the physical aspects of martingale is still needed. 

\acknowledgements
The author thanks Juan Parrondo for his questions during the {\it kT}\,Log2'24 conference. The author thanks Muhittin Mungan for reading the draft.

%%%%%%%%%%%%%%%%%%%%%%%%%%%%%%%%
\bibliographystyle{apsrev4-2.bst}
\bibliography{ken_LNP_sar.bib}

\end{document}